\documentclass[ALICE,manyauthors]{cernphprep}
\usepackage{graphicx}
\usepackage{amssymb}
\usepackage{color}
\usepackage{hyperref}
\usepackage{makeidx}
\usepackage{ulem}

\newcommand{\pT}{\ensuremath{p_{\rm t}}} 
\newcommand{\mT}{\ensuremath{m_{\rm t}}} 
 
\newcommand{\dedx}{d$E$/d$x$}

\newcommand{\pp}{pp}
\newcommand{\s}{$\sqrt{s}$} 

\newif\ifcomment
\newif\ifcolordiff
\newcommand{\orig}[1]      {}
\newcommand{\repl}[1]      {}
\ifcolordiff
\renewcommand{\orig}[1]      {\protect\textcolor{red}{\sout{#1}}}
\renewcommand{\repl}[1]      {\protect\textcolor{blue}{#1}}
\else
\renewcommand{\repl}[1]      {#1}
\fi

\begin{document}%
%
%
\begin{titlepage}
\PHnumber{2012-001}                 
\PHdate{28 May 2012}              
%
%
\title{Neutral pion and $\eta$ meson production \\
  in proton-proton collisions at
  \s~=~0.9~TeV and \s~=~7~TeV}
\ShortTitle{$\pi^0$ and $\eta$ spectra}   
%
\Collaboration{ALICE Collaboration%
         \thanks{See Appendix~\ref{app:collab} for the list of collaboration
                      members}}
\ShortAuthor{ALICE Collaboration}      
\begin{abstract}
  The first measurements of the invariant differential cross sections of
  inclusive $\pi^0$ and $\eta$ meson production at mid-rapidity in
  proton-proton collisions at \s~=~0.9~TeV and \s~=~7~TeV are reported. The $\pi^0$ measurement covers the ranges
  $0.4<\pT<7$ GeV/$c$ and $0.3<\pT<25$ GeV/$c$ for these two energies,
  respectively. The production of $\eta$ mesons was measured at \s~=~7~TeV in the range $0.4<\pT<15$~GeV/$c$. Next-to-Leading
  Order perturbative QCD calculations, which are
  consistent with the $\pi^0$ spectrum at \s~=~0.9~TeV, 
  overestimate those of $\pi^0$ and $\eta$ mesons at \s~=~7~TeV,  
but agree with the measured $\eta/\pi^0$ ratio 
at \s~=~7~ TeV.
\end{abstract}
\end{titlepage}
\setcounter{page}{2}

\section{Introduction}

Hadron production measurements in proton-proton collisions at 
the Large Hadron Collider (LHC)~\cite{Evans:2008zzb}
energies open a new, previously unexplored domain in particle phy\-sics,
which allows validation of the predictive power of Quantum
Chromo Dynamics (QCD) \cite{Gross:1973ju}. A quantitative description of hard
processes is provided by perturbative QCD (pQCD) supplemented with
parton distribution functions (PDF) $f(x)$ and fragmentation functions
(FF) $D(z)$, where $x$ is the fraction of the proton longitudinal
momentum carried by a parton and $z$ is the ratio of the observed hadron
momentum to the final-state parton momentum.  
Due to the higher collision energy at the LHC, the PDF and FF can
be probed at lower values of $x$ and $z$, respectively, than in
previous experiments. Such measurements can provide further
constraints on these functions, which are crucial for pQCD predictions for LHC energies.
Furthermore,  while pion production at the Relativistic Heavy Ion Collider 
 (RHIC) \cite{Reardon:1988hg} is considered to be dominated by gluon fragmentation 
 only for $\pT < 5-8$~GeV/$c$ \cite{Vogt,Adare:2010cy}, at
 LHC energies it should remain dominant for $\pT < 100$~GeV/$c$
 \cite{Sassot:2010bh,Chiappetta:1992uh}. Theoretical estimates
\cite{Sassot:2010bh} suggest that the fraction of pions 
originating from gluon fragmentation remains above 75~\% in the $\pT$
range up to 30~GeV/$c$. Here, the measurement of the
$\pi^0$ production cross section at LHC energies provides constraints
on the gluon to pion fragmentation \cite{deFlorian:2007aj} in a new energy
regime. In addition, the strange quark content of the $\eta$ meson makes the
comparison to pQCD relevant for possible differences of
fragmentation functions with and without strange quarks
\cite{Aidala:2010bn}. Furthermore, the precise measurement of $\pi^0$ and
$\eta$ meson spectra over a large $\pT$ range is a
prerequisite for understanding the decay photon (electron) background for
a direct photon (charm and beauty) measurement.  Finally, a significant
fraction of hadrons at low $\pT$ is produced in \pp ~collisions via
soft parton interactions, which cannot be well described within the
framework of pQCD. In this kinematic region commonly used event generators like
 PYTHIA~\cite{Sjostrand:2006za} or PHOJET~\cite{Engel:1995sb} have to resort to
phenomenological models tuned to available experimental
data delivered by lower-energy colliders like Sp$\bar{\mbox{p}}$S,
RHIC, and  Tevatron~\cite{Aaltonen:2009ne}, to adequately describe hadron production.
The large increase in center-of-mass energy at the 
LHC provides the possibility for a stringent test of the 
extrapolations based on these models.

This paper presents the first measurement of neutral pion and $\eta$
meson production in proton-proton collisions at center-of-mass
energies of $\sqrt{s}=0.9$~TeV and 7~TeV in a wide $\pT$ range with
the ALICE detector~\cite{Aamodt:2008zz}.  The paper is organized as
follows: description of the subdetectors used for these measurements,
followed by the details about the data sample, as well as about event
selection and photon identification, is given in
section~\ref{sec:Detector}. Section~\ref{sec:Reconstruction} describes
the algorithms of neutral meson extraction, methods of production
spectra measurement, and shows the systematic uncertainty
estimation. Results and their comparison with pQCD calculations are
given in section~\ref{sec:Results}.

\section{Detector description and event selection}
\label{sec:Detector}

Neutral pions and $\eta$ mesons are measured in ALICE via the
two-photon decay channel.  The photons are detected with two methods
in two independent subsystems, with the Photon Spectrometer (PHOS)
\cite{PHOS_TDR} and with the photon conversion method (PCM) in the
central tracking system employing the Inner Tracking System (ITS)
\cite{Aamodt:2010ys} and the Time Projection Chamber (TPC)
\cite{Alme:2010ke}.  The latter reconstructs and identifies photons
converted to $\rm{e}^+\rm{e}^-$ pairs in the material of the inner
detectors. The simultaneous measurements with both methods with
completely different systematic uncertainties and with momentum
resolutions having opposite dependence on momentum provide a
consistency check of the final result.

The PHOS detector consists at present of three modules installed at a
distance of 4.60~m from the interaction point. PHOS covers the
acceptance of $260^\circ<\varphi<320^\circ$ in azimuthal angle and
$|\eta|<0.13$ in pseudorapidity. Each module has 3584 detection
channels in a matrix of $64\times56$ cells. Each detection channel
consists of a lead tungstate, $\mbox{PbWO}_4$, crystal of $2.2\times
2.2$~cm$^2$ cross section and 18~cm length, coupled to an avalanche
photo diode and a low-noise charge-sensitive preamplifier. PHOS
operates at a temperature of $-25~^\circ$C at which the light yield of
the $\mbox{PbWO}_4$ crystal is increased by about a factor 3 compared
to room temperature. PHOS was calibrated in-situ by equalizing mean
deposited energies in each channel using events with pp collisions.

The Inner Tracking System (ITS)  \cite{Aamodt:2008zz} consists of six layers
equipped with Silicon Pixel Detectors (SPD) positioned at a radial 
distance of 3.9 cm and 7.6 cm, Silicon Drift Detectors (SDD) at 15.0
cm and 23.9 cm, and Silicon Strip Detectors (SSD) at 38.0 cm and 43.0 cm. The two
innermost layers cover a pseudorapidity range of $|\eta|<2$ and
$|\eta| < 1.4$, respectively. 

The Time Projection Chamber (TPC)
\cite{Alme:2010ke} is a large (85~m$^3$) cylindrical drift detector filled
with a Ne/CO$_2$/N$_2$ (85.7/9.5/4.8\%) gas mixture. It is the main
tracking system of the Central Barrel.  For the maximum track length
of 159 clusters it covers a pseudorapidity range of $|\eta|<0.9$ over
the full azimuthal angle.  In addition, it provides particle
identification via the measurement of the specific ionisation energy loss
(d$E$/d$x$) with a resolution of 5.5\% \cite{Alme:2010ke}. The ITS and the
TPC are aligned with respect to each other to the level of few hundred
$\mu$m  using cosmic-ray and proton-proton collision data \cite{Aamodt:2010ys}.

The event selection was performed with the VZERO
detector \cite{VZERO} in addition to the SPD. The VZERO is a forward scintillator hodoscope with two
segmented counters located at $3.3$~m and $-0.9$~m from the
interaction point. They cover the pseudorapidity ranges $2.8<\eta<5.1$
and $-3.7<\eta<-1.7$, respectively.

The proton-proton collision data used in this analysis were collected
by the ALICE experiment in 2010 with the minimum bias trigger MB$_{\rm
 OR}$~\cite{Aamodt:2010pp}. This trigger required the crossing of two
filled bunches and a signal in at least one of the two SPD pixel
layers or in one of the VZERO counters.  An offline selection based on
time and amplitude signals of the VZERO detectors and the SPD was
applied to reject beam-induced and noise background
\cite{Aamodt:2010pp}. \repl{Pileup collision events were identified imposing
a criterion based on multiple primary vertices reconstructed with the SPD detector,
and removed from the further analysis.}
The cross sections for the MB$_{\rm OR}$
trigger have been calculated from other measured cross sections at the
same energies with appropriate scaling factors.  At $\sqrt{s}=7$~TeV
the cross section for the coincidence between signals in the two VZERO
detectors, $\sigma_{\rm{MB_{\rm AND}}}$, was measured in a
Van-der-Meer scan~\cite{alice_sigma7TeV}, and the relative factor
$\sigma_{\rm MB_{\rm AND}}$/$\sigma_{\rm MB_{\rm OR}}=0.87$\orig{$2$}\repl{$3$}
\orig{$\pm 0.003$}\repl{with negligible error} as obtained from data was used.  
At $\sqrt{s}=0.9$~TeV the cross
section $\sigma_{\rm MB_{\rm OR}}$ has been calculated from the
inelastic cross section measured in $\rm{p}\bar{\rm{p}}$ collisions at
$\sqrt{s}=0.9$~TeV~\cite{Alner:1986iy} and relative factor
$\sigma_{\rm MB_{OR}}/\sigma_{\rm inel}=0.91$\orig{$6 \pm 0.013$}\repl{$^{+0.03}_{-0.01}$} 
estimated from Monte Carlo simulations~\orig{[21]}\repl{\cite{alice_sigma7TeV}}.

Table~\ref{table-cs} shows the values of the cross section obtained at
both energies as well as the integrated luminosity of the total data
samples used.  In the photon conversion analysis, only events with a
reconstructed vertex ($\sim$90\% of the total) are inspected, and
those events with a longitudinal distance (i.e. along the beam
direction) between the position of the primary vertex and the
geometrical center of the apparatus larger than 10~cm are discarded.
The analysis using PHOS as well as Monte Carlo simulations show that
the number of $\pi^0$s in events without a reconstructed vertex is
below 1\% of the total number of $\pi^0$s.

\begin{table}[b]
\begin{center}
\begin{tabular}{|c|c|c|}
\hline
\s (TeV) & $\sigma_{\rm MB_{\rm OR}}$ (mb) & $\sigma_{pp}^{\rm INEL}$ (mb) \\ \hline
$0.9$ & \orig{$46.1 \pm 1.1$}\repl{$47.8^{+2.4}_{-1.9}$}(syst) & \orig{$50.3 \pm 0.4(\rm stat) \pm 1.1$}\repl{$52.5 \pm 2$}(syst)\\ \hline
$7$ & $62.$\orig{$5$}\repl{$2$}$ \pm 2.2$(syst) & $73.2$ \orig{$\pm 1.1^{\rm model}$}\repl{$^{+2.0}_{-4.6}$} $\pm 2.6^{\rm lumi}$ \\ \hline
\end{tabular}\\[2pt]
\begin{tabular}{|c|c|c|c|}
\hline
\s (TeV) &\multicolumn{3}{|c|}{  \rule{54pt}{0pt} $\cal{L}$ $({\rm nb}^{-1})$  \rule{54pt}{0pt} } \\
\cline{2-4}
&  \rule{8pt}{0pt} PCM \rule{7pt}{0pt}  &  \rule{3pt}{0pt} PHOS $\pi^0$ \rule{2pt}{0pt}  & PHOS $\eta$ \\ \hline
0.9 & 0.14 & 0.14 & \\ \hline
7 & 5.6 & 4.0 & 5.7 \\ \hline
\end{tabular}
\end{center}
\caption{Cross sections of the reactions and integrated luminosities of the measured data samples for the two beam energies (top), 
and luminosities used in the different analyses for the 7 TeV data (bottom).}
\label{table-cs}
\end{table}%

To maximize the pion reconstruction efficiency in PHOS, only
relatively loose cuts on the clusters (group of crystals with deposited energy and common edges) were used: 
the cluster energy was required to be above the minimum ionizing 
energy $E_{\rm  cluster}>0.3$~GeV and the minimum number of crystals in a 
cluster was three to reduce the contribution of non-photon clusters.

Candidate track pairs for photon conversions were reconstructed using a secondary
vertex (V0) finding  algorithm \cite{Alessandro:2006yt}. 
In order to select photons among all secondary vertices (mainly
$\gamma$, K$^0_S$, $\Lambda$ and $\bar{\Lambda}$), electron selection
and pion rejection cuts were applied. The main particle identification
(PID) selection used the specific energy loss in the TPC (\dedx). 
The measured \dedx\ of electrons was required to lie in the interval
$[-4\sigma_{{\rm d}E/{\rm d}x},+5\sigma_{{\rm d}E/{\rm d}x}]$ around the expected value. In
addition, pion contamination was further reduced by a cut of 
$2\sigma$ above the nominal pion \dedx\ in the momentum range
of $0.25$~GeV/$c$ to $3.5$~GeV/$c$ and a cut of 0.5$\sigma$
at higher momenta. 
 For the $\gamma$ reconstruction constraints on the
reconstructed photon mass and on
the opening angle between the reconstructed photon momentum vector and
the vector joining the collision vertex and the conversion point were
applied. These constraints were implemented as a cut on the $\chi^2(\gamma)$ defined using a
reconstruction package for fitting decay particles \cite{alikf}. 
The photon measurement contains information on the direction which allows 
to reduce the contamination from secondary neutral pions. 
With the conversion method a precise $\gamma$-ray tomograph of the
ALICE experiment has been obtained \cite{ALICE_X0}.  The integrated
material budget for $r < 180$~cm and $|\eta|<$0.9 is
11.4\orig{$^{+0.39}_{-0.71}$}\repl{$\pm0.5$}\% $X_0$ as extracted from detailed comparisons
between the measured thickness and its implementation in Monte Carlo
simulations based on the GEANT 3.21 package\orig{.}\repl{ using the same simulation runs
for the material studies as for the $\pi^0$ measurement.}
Photon pairs with an opening angle larger than 5~mrad were selected for the meson analysis.

\section{Neutral meson reconstruction}
\label{sec:Reconstruction}

\begin{figure}[htb]
  \hfil
  \includegraphics[width=0.45\textwidth]{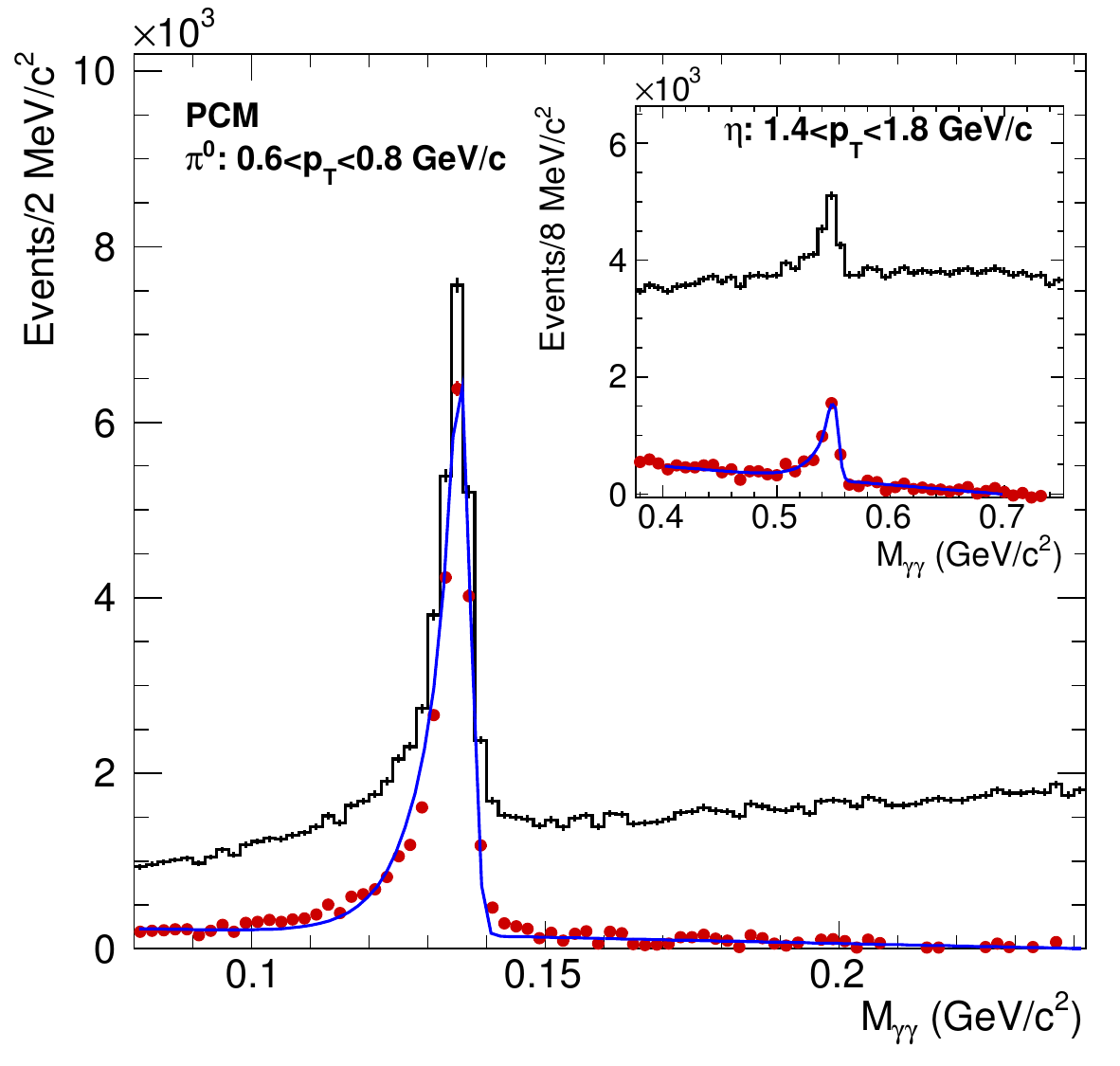}
  \hfil
  \includegraphics[width=0.45\textwidth]{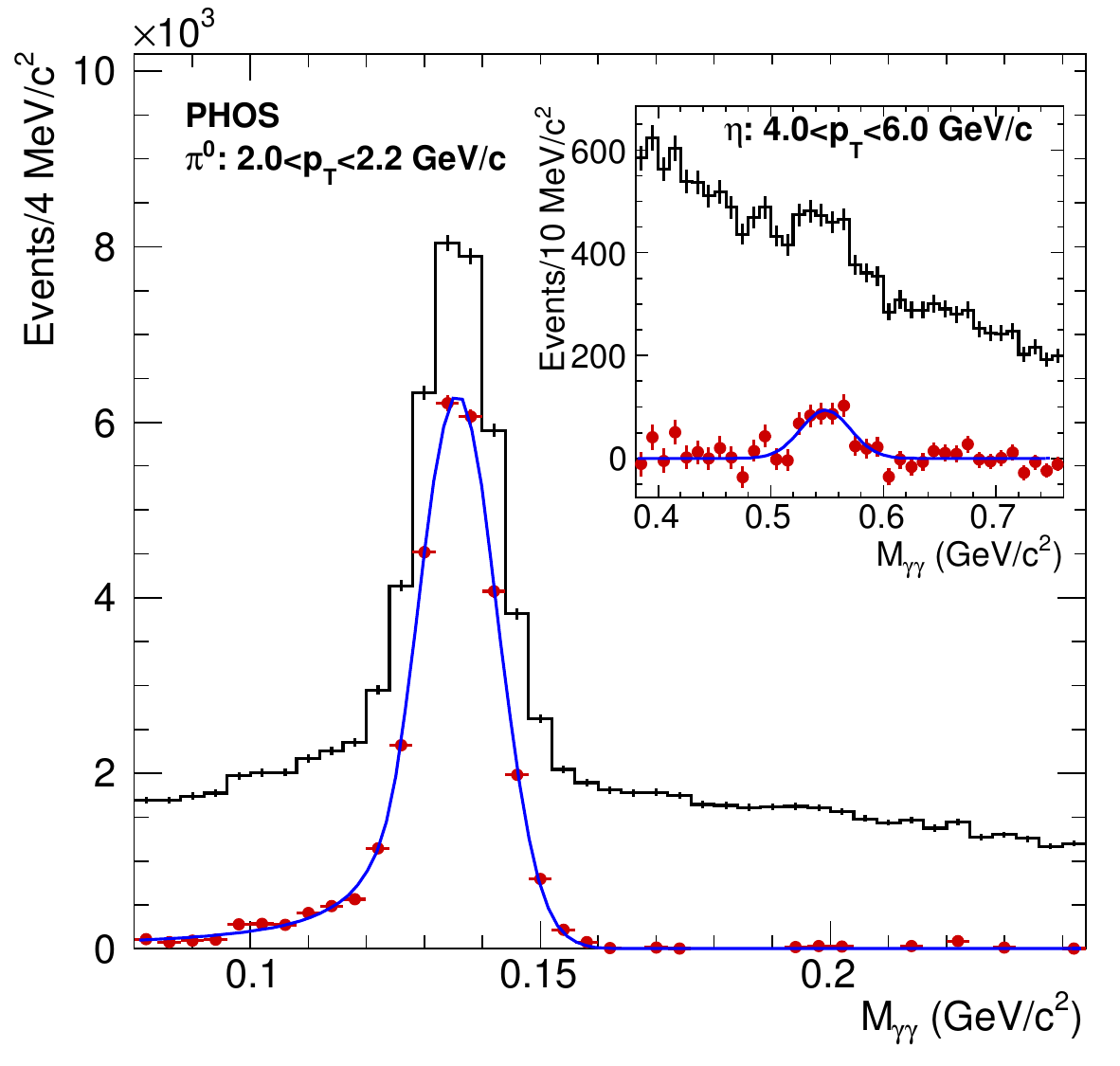}
  \hfil
  \caption{Invariant mass spectra in selected \pT\ slices in PCM
    (left) and PHOS (right) in the $\pi^0$ and $\eta$ meson mass
    regions. The histogram and the bullets show the data before and
    after background subtraction, respectively. The curve is a
    fit to the invariant mass spectrum after background subtraction.}
  \label{fig:InvMassSpec}
\end{figure}

\begin{figure}[htb]
  \centering
  \includegraphics[width=0.48\textwidth]{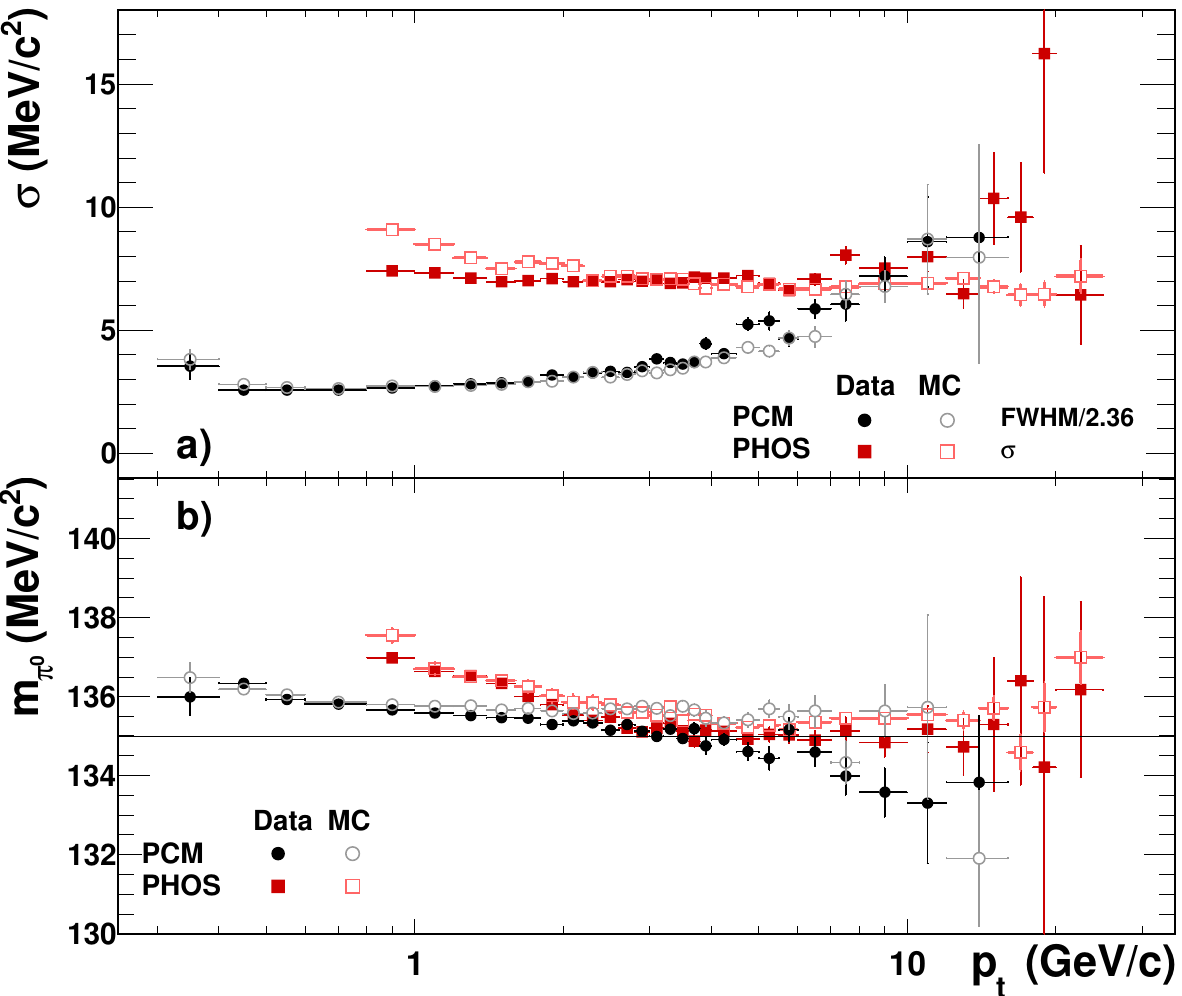}
  \caption{Reconstructed $\pi^0$ peak width (a) and position (b)
    as a function
    of $\pT$ in \pp ~collisions at $\sqrt{s}=7$~TeV
    in PHOS and in the photon
    conversion method (PCM) compared to Monte Carlo simulations. 
    The horizontal line in (b) indicates the nominal $\pi^0$
    mass.}
  \label{fig:InvMass}
\end{figure}

Neutral pions and $\eta$ mesons are reconstructed as excess yields,
visible as peaks at their respective rest mass, above the
combinatorial background in the two-photon invariant mass
spectrum. Invariant mass spectra demonstrating the $\pi^0$ and
$\eta$ mesons peak in some selected \pT\ slices are
shown in Fig.\ref{fig:InvMassSpec} by the histogram. 
The background is determined by mixing photon pairs from different
events and is normalized to the same event background at the right
side of the meson peaks. A residual correlated background is further
subtracted using a linear or second order polynomial fit. The
invariant mass spectrum after background subtraction, depicted by
bullets in Fig.\ref{fig:InvMassSpec}, was fitted to obtain the
$\pi^0$ and $\eta$ peak parameters (a curve). The number
of reconstructed $\pi^0$s ($\eta$s) is obtained in each $\pT$ bin by
integrating the background subtracted peak within 3 standard
deviations around the mean value of the $\pi^0$ ($\eta$) peak position
in the case of PHOS.  In the PCM measurement the integration windows
were chosen to be asymmetric ($m_{\pi^0}$-0.035~GeV/$c^2$,
$m_{\pi^0}$+0.010~GeV/$c^2$) and ($m_{\eta}$-0.047~GeV/$c^2$,
$m_{\eta}$+0.023~GeV/$c^2$) to take into account the left side tail of
the meson peaks due to bremsstrahlung. For the same reason in the case
of PCM the full width at half maximum (FWHM) instead of the Gaussian
width of the peak was used.  We vary the normalization and integration
windows to estimate the related systematic uncertainties.  The peak
position and width from the two analyses compared to Monte Carlo
simulations are shown in Fig.~\ref{fig:InvMass} as a function of
$\pT$.

The reconstruction efficiency $\epsilon $ and acceptance $A$ are
calculated 
in Monte Carlo simulations tuned to reproduce the detector
response. In the PHOS case, the tuning included a
4.5\% energy non-linearity observed in real data at $E<1$~GeV and not 
reproduced by the GEANT simulations and an additional 6\% 
channel-by-channel decalibration. 
In the PCM case, an additional smearing in 
each momentum component given by $\sigma=\sqrt{\sigma_0^2+\sigma_1^2\cdot p^2}$ 
with $\sigma_0=0.011$~GeV/$c$ and $\sigma_1=0.007$ was necessary to reproduce the measured
width of the $\pi^0$ peak.
 PYTHIA~\cite{Sjostrand:2006za} and PHOJET~\cite{Engel:1995sb} event
generators and single particle simulations were used as input.  
The small photon conversion
probability of about 8.5\%, compensated by the large TPC acceptance,
translates into $\epsilon\cdot A$ of about
$2\times10^{-3}$ at $\pT > 1$~GeV/$c$
and decreases at lower $\pT$ due to the decrease of the efficiency of soft
electron reconstruction and conversion probability.  In the PHOS case,
 the acceptance $A$ is zero for $\pT<0.4$~GeV/$c$, 
$\epsilon\cdot A$ increases with $\pT$ and
saturates at about $2.0\times 10^{-2}$ at $\pT>15$~GeV/$c$.  At high
$\pT>25$~GeV/$c$ the efficiency decreases due to cluster merging.

The invariant differential cross section of $\pi^0$ and $\eta$ meson
production were calculated as
\begin{equation}
  E \frac{{\rm d}^3 \sigma}{{\rm d}p^3} =
  \frac{1}{2\pi} \frac{\sigma_{\rm MB_{\rm OR}}}{N_{\rm events}} \frac{1}{\pT} 
  \frac{1}{\epsilon \,A\,Br}\frac{N^{\pi^0 (\eta)}}{\Delta y \Delta \pT}\,,
\end{equation}
where $\sigma_{\rm {MB_{\rm OR}}}$ is the interaction cross section
for the MB$_{\rm OR}$ trigger for \pp ~collisions at $\sqrt{s}=0.9$~TeV
or $\sqrt{s}=7$~TeV, $N_{\rm events}$ is the number of MB$_{\rm OR}$
events, $\pT$ is the transverse momentum within the bin to which 
the cross section has been assigned after the correction for the finite bin width $\Delta \pT$ (see below),
$Br$ is the branching ratio of the $\pi^0$ ($\eta$) meson to the two
$\gamma$ decay channel and $N^{\pi^0(\eta)}$ is the number of
reconstructed $\pi^0$ ($\eta$) mesons in a given $\Delta y$ and
$\Delta \pT$ bin. Finally, the invariant cross sections were corrected
for the finite $\pT$ bin width following the prescription in
\cite{Lafferty:1994cj}, keeping the $y$ values equal to the bin averages and 
calculating the $\pT$ position at which the differential cross section coincides with
the bin average. The Tsallis fit (see below) was used for the correction.
Secondary $\pi^0$'s from weak decays or hadronic interactions in the detector
material are subtracted using Monte Carlo simulations. The contribution 
from K$^0_S$ decays is scaled using the measured K$^0_S$ spectrum at 
$\sqrt{s}=0.9$~TeV \cite{Aamodt:2011zza} or the charged kaon spectra 
at $\sqrt{s}=7$~TeV \cite{Floris:2011ru}.
The measured $\pi^0$ and $\eta$ meson spectra at the center-of-mass
energy of $\sqrt{s}=7$~TeV cover a $\pT$ range from 0.3 to 25~GeV/$c$ and 
from 0.4 to 15~GeV/$c$, respectively; the $\pi^0$ spectra at 
$\sqrt{s}=0.9$~TeV cover a $\pT$ range  from 0.4 to 7~GeV/$c$.

\begin{table}[ht]
  \centering
  \begin{tabular}{|c|c|c|c|c|} 
   \hline
                          &  \multicolumn{2}{|c|}{PHOS}   \\
   \cline{2-3}
                          & $\pT=1.1$~GeV/$c$ &  $\pT=7.5$~GeV/$c$   \\
    \hline
    Yield extr.           &   $\pm$2.1   &  $\pm$2.5      \\
    Non-linearity         &   $\pm$9.0   &  $\pm$1.5      \\
    Conversion            &   $\pm$3.5   &  $\pm$3.5      \\
    Absolute energy scale &   $\pm$0.7   &  $\pm$1.0      \\
    Acceptance            &   $\pm$1.0   &  $\pm$1.0      \\
    \repl{Calibration and alignment} &   \repl{$\pm$7.0}   &  \repl{$\pm$3.0}  \\
    \repl{Pileup}                    &   \repl{$\pm$0.8}   &  \repl{$\pm$0.8}  \\
    \hline 
    Total                 &   $\pm$\orig{10.0}\repl{12.5}\%  &  $\pm$\orig{4.8}\repl{6.0}\%  \\
   \hline
   \hline
                              &  \multicolumn{2}{|c|}{PCM}  \\
   \cline{2-3}
                              & $\pT$=1.1~GeV/$c$ &  $\pT$=7.5~GeV/$c$   \\
    \hline
    Material Budget           &  \orig{$-8.8~,~+4.8$}\repl{$\pm 9.0$}    & \orig{$-8.8, +4.8$}\repl{$\pm 9.0$}    \\
    Yield extraction          &  \orig{$-1.2~,~+0.2$}\repl{$\pm 0.6$}    & \orig{$-9.3, +9.5$}\repl{$\pm 4.9$}    \\
    PID                       &  \orig{$-0.15,~+0.1$}\repl{$\pm 0.1$}    & \orig{$-8.9, +1.7$}\repl{$\pm 5.4$}    \\
    $\chi^2 (\gamma)$         &  \orig{$-0.6~,~+0.1$}\repl{$\pm 0.3$}    & \orig{$-7.7, +4.7$}\repl{$\pm 6.2$}    \\
    Reconstruction $\epsilon$ &  \orig{$-0.4~,~+3.8$}\repl{$\pm 1.9$}    & \orig{$-9.8, +7.8$}\repl{$\pm 4.9$}    \\
    \hline 
    Total                     &  \orig{$-8.9\%,~+6.1$}\repl{$\pm 9.2$}\%   & \orig{$-20\%,~+14$}\repl{$\pm 14.0$}\%\\
    \hline
  \end{tabular}
  \caption{Summary of the relative systematic errors for the PHOS and the PCM analyses. }
  \label{tab:SysErrs}
\end{table}

A summary of the systematic uncertainties is shown in Table\ \ref{tab:SysErrs} for 
two different $\pT$ values.
In PHOS, the significant source of systematic errors both at low and 
high $\pT$ is the raw yield extraction. It was estimated by varying 
the fitting range and the assumption about the shape of the background around the peak. 
The uncertainty related to the non-linearity of PHOS which dominates
at low $\pT$ was estimated by introducing different non-linearities into 
the MC simulations under the condition that the simulated $\pT$-dependence 
of the $\pi^0$ peak position and peak width is still consistent with data.
\repl{The uncertainties on the calibration and alignment were estimated in Monte Carlo simulations 
by varying the calibration parameters and the relative module positions within the expected tolerances.
The uncertainty related to the pileup event rejection was evaluated in data by estimating the fraction of 
unidentified pileup events by extrapolating the distance and the number of contributing tracks of found 
pileup vertices to zero.}
The uncertainty of the conversion probability was estimated comparing 
measurements without magnetic field to the standard measurements with 
magnetic field. 
In the measurements with converted photons, the main sources of systematic errors are the
knowledge of the material budget (dominant at low $\pT$), raw yield extraction, PID, 
the photon $\chi^2$ cut and reconstruction efficiency. The contribution from the raw yield 
extraction was estimated by changing the normalization range, the integration window,
and the combinatorial background evaluation. The PID, photon $\chi^2(\gamma)$ cut and 
reconstruction efficiency was estimated by evaluating stability of the results 
after changing the cut values.

\section{Results and comparison with pQCD}
\label{sec:Results}

\begin{figure}[htp]
  \centering
  \includegraphics[width=0.55\textwidth]{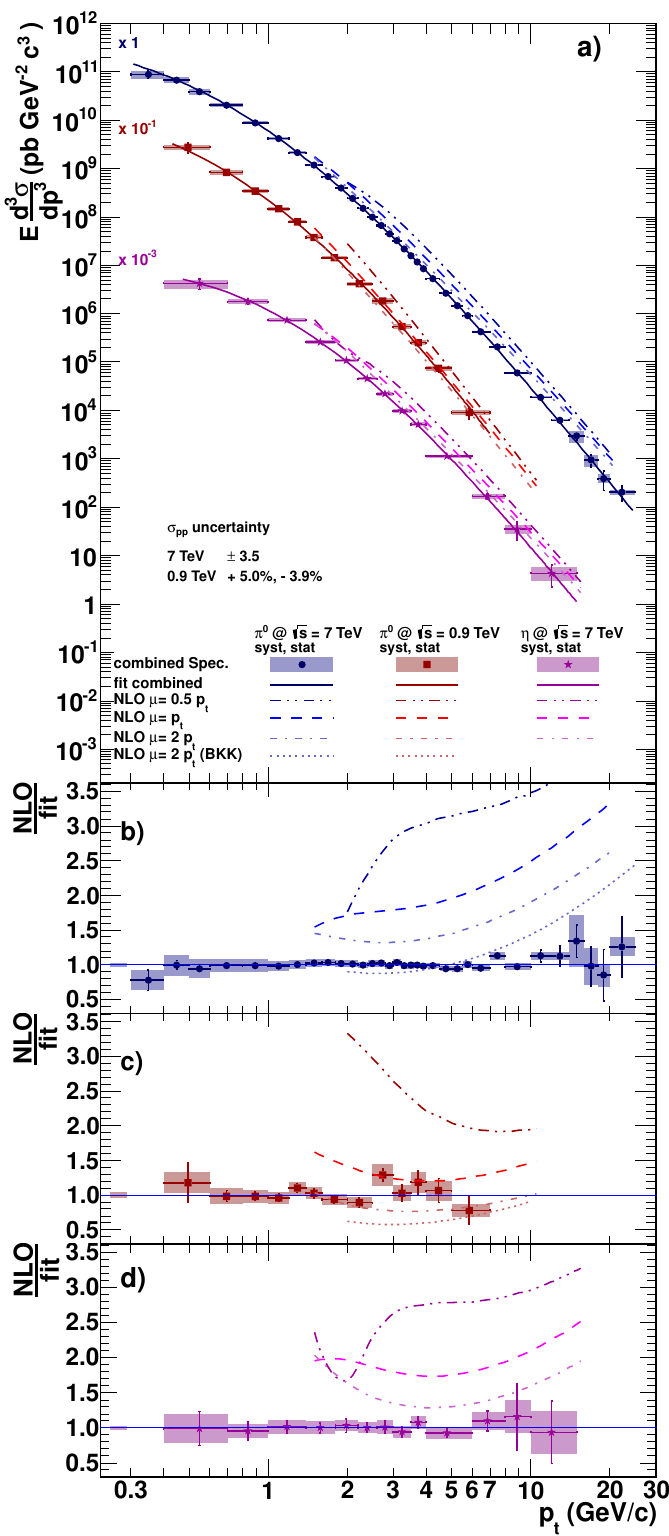}
  \caption{a) Differential invariant cross section of $\pi^0$
    production in \pp ~collisions at $\sqrt{s}=7$~TeV (circles) and
    0.9~TeV (squares) and of $\eta$ meson production at
    $\sqrt{s}=7$~TeV (stars). The lines and the boxes represent the
    statistical and systematic error of the combined measurement respectively.
    The uncertainty on the \pp ~cross section is not included.
    NLO pQCD calculations using
    the CTEQ6M5 PDF and the DSS (AESS for $\eta$ mesons) FF for three scales $\mu=0.5\pT$,
    $1\pT$ and $2\pT$ are shown.
    Dotted lines in panels b) and
    c) correspond to the ratios using the BKK FF. Ratio of the NLO
    calculations to the data parametrisations are shown in panels b),
    c) and d). The full boxes represent the uncertainty on the \pp~
    cross sections.}
\label{fig:CrossSection}
\end{figure}

The combined spectrum is calculated as a weighted average using
statistical and systematic errors of the individual analyses
\cite{Nakamura:2010zzi}.  The combined production cross sections are
shown in Fig.~\ref{fig:CrossSection} a).
The combined spectra \repl{including statistical and systematic errors} are fitted with the Tsallis function \cite{Tsallis:1987eu} 
\begin{equation}
   \displaystyle
\!\!\!\!\!\!\!\!\!\!\!\!\!\!\!
    E \frac{{\rm d}^3 \sigma}{{\rm d}p^3} = 
    \displaystyle
    \frac{\sigma_{pp}^{\rm INEL}}{2\pi}A
    \frac{c \cdot (n-1)(n-2)}{nC\left[ nC+m(n-2)\right]}  
    \displaystyle\left(1+\frac{\mT-m}{nC}\right)^{-n},
    \label{eq:Tsallis}
\end{equation}
where the fit parameters are $A$, $C$ and $n$, $\sigma_{\rm pp}$ is
the proton-proton inelastic cross section, $m$ is the meson rest mass
and $\mT=\sqrt{m^2+\pT^2}$ is the transverse mass. The fit parameters
are shown in Table~\ref{lab:TsallisParam}\orig{, where the uncertainties are
the quadratic sum of the statistical and systematical
uncertainties}. The property of the Tsallis function (\ref{eq:Tsallis})
is such that the parameter $A$ is equal to the integral of this
function over $\pT$ from 0 to infinity, $A={\rm d}N/{\rm d}y$, and
thus can be used as an estimation of the total yield at $y=0$\orig{.} 
\repl{per inelastic pp collision.} The additional uncertainty on the parameter 
$A$ due to the spectra normalization of \orig{$1.4$}\repl{$^{+3.2}_{-1.1}$}\% and 
\orig{$2.8$}\repl{$^{+7.0}_{-3.5}$}\% at $\sqrt{s}=900$~GeV, 
and $7$~TeV respectively\orig{, which is not shown in the table}\repl{, is not included}.
The found parameters of the Tsallis function for $\pi^0$ production spectrum in
pp collisions at $\sqrt{s}=900$~GeV are in agreement with those for
the $\pi^+ + \pi^-$ spectra measured by the ALICE collaboration at the
same energy \cite{Aamodt:2011zj}.

\begin{figure}[htp]
  \centering
  \includegraphics[width=0.48\textwidth]{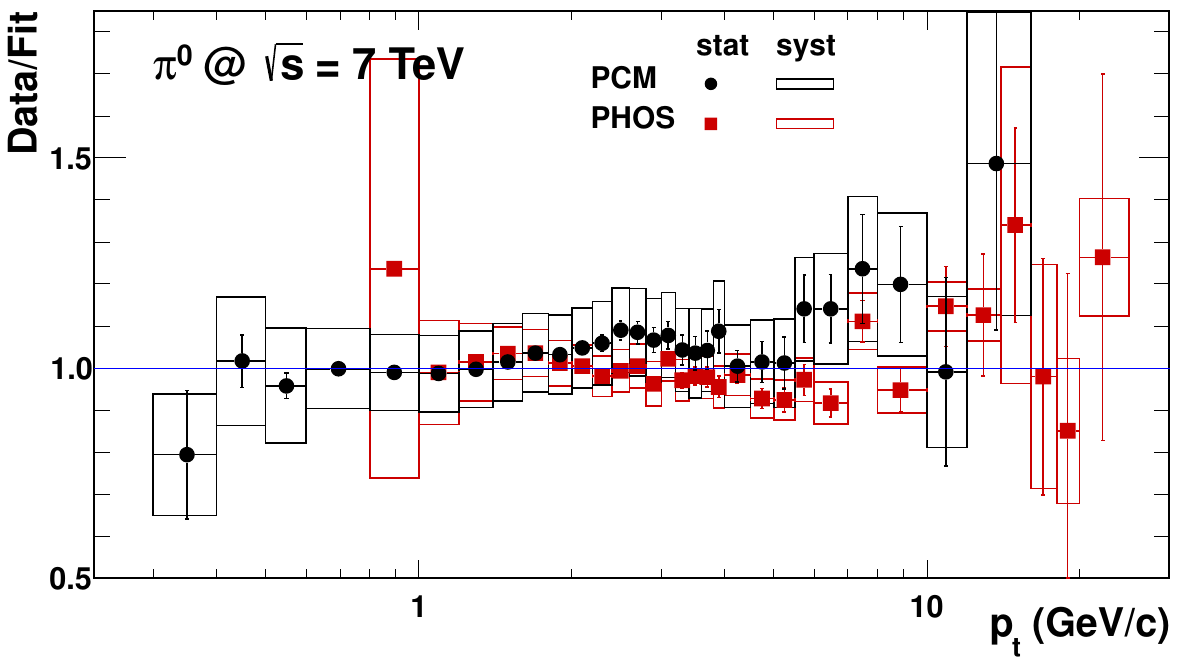}
  \caption{Ratio of the two independent $\pi^0$ meson measurements
    to the fit of the combined normalized invariant production cross
    section of $\pi^0$ mesons in \pp ~collisions at
    $\sqrt{s}=7$~TeV.}
  \label{fig:YieldNormalized}
\end{figure}

\begin{table}[t]
  \centering
  \begin{tabular}{|c|c|c|c|c|} \hline
    Meson & \s         & $A$       & $C$   & $n$ \\
          &    TeV     &                   & (MeV/$c^2$) &     \\\hline
    $\pi^0$& 0.9 & $1.5 \pm 0.3$   & $132 \pm 15$ & $7.8 \pm 0.5$ \\\hline
    $\pi^0$& 7   & \orig{$2.45 \pm 0.07$}\repl{$2.40 \pm 0.15$} & \orig{$140$}\repl{$139$}$ \pm 4$ & \orig{$6.90$}\repl{$6.88$}$ \pm 0.07$ \\\hline
    $\eta$ & 7   & \orig{$0.22$}\repl{$0.21$}$ \pm 0.03$ & $229 \pm 21$ & \orig{$6.9$}\repl{$7.0$}$ \pm 0.5$ \\\hline
  \end{tabular}
  \caption{Fit parameters of the Tsallis parametrisation
    (\ref{eq:Tsallis}) to the combined invariant production yields of
    $\pi^0$ and $\eta$ mesons for inelastic events. \orig{The errors are
    statistical and systematic added in quadrature.}
    The uncertainty on the parameter $A$ due to the spectra normalization 
    of \orig{$1.4$}\repl{$^{+3.2}_{-1.1}$}\% and
    \orig{$2.8$}\repl{$^{+7.0}_{-3.5}$}\% at $\sqrt{s}=900$~GeV,
    and $7$~TeV respectively, is not included.}
  \label{lab:TsallisParam}
\end{table}
The ratio of the data points of the two methods to the combined fit,
shown in Fig.~\ref{fig:YieldNormalized}, illustrates the consistency
between the two measurements.

We compare our results with Next-to-Leading Order~(NLO) pQCD calculations
using the PDF CTEQ6M5 and DSS $\pi^0$ \cite{deFlorian:2007aj}, BKK
$\pi^0$ \cite{Binnewies:1994ju} and AESSS $\eta$ \cite{Aidala:2010bn}
NLO fragmentation functions, see Fig.~\ref{fig:CrossSection} a). The
data and NLO predictions are compared via a ratio with the fit to the
measured cross section.  This is shown in the bottom panels (b), (c)
and (d) in Fig.~\ref{fig:CrossSection}. In the NLO calculations the
factorization, renormalization and fragmentation scales are chosen to
have the same value given by $\mu$.  The uncertainty in the inelastic
\pp\ cross section is represented by the full boxes at unity. At
$\sqrt{s}=0.9$~TeV the NLO calculations at $\mu=1\,\pT$ describe the
measured $\pi^0$ data well, while at $\sqrt{s}=7$~TeV the higher scale
($\mu = 2\, \pT$) and a different set of fragmentation functions are
required for a description of the data. However, the latter parameter
set does not provide a good description of the low energy data. In any
case, the NLO pQCD calculations show a harder slope compared to the
measured results. Using the INCNLO program~\cite{Aurenche:1999nz}, we
tested different parton distribution functions (CTEQ5M, CTEQ6M, MRS99)
and different fragmentation functions (BKK, KKP, DSS) and found a
similar result: pQCD predicts harder slopes, and variation of PDFs and
FFs does not change the shape, but results mainly in the variation of
the absolute cross section.  A similar trend is observed for the
$\eta$ meson (a higher scale $\mu = 2 \pT$ is required), although the
discrepancy is less significant due to the larger error bars and
smaller $\pT$ reach.

\begin{figure}[hbtp]
  \centering
  \includegraphics[width=0.48\textwidth]{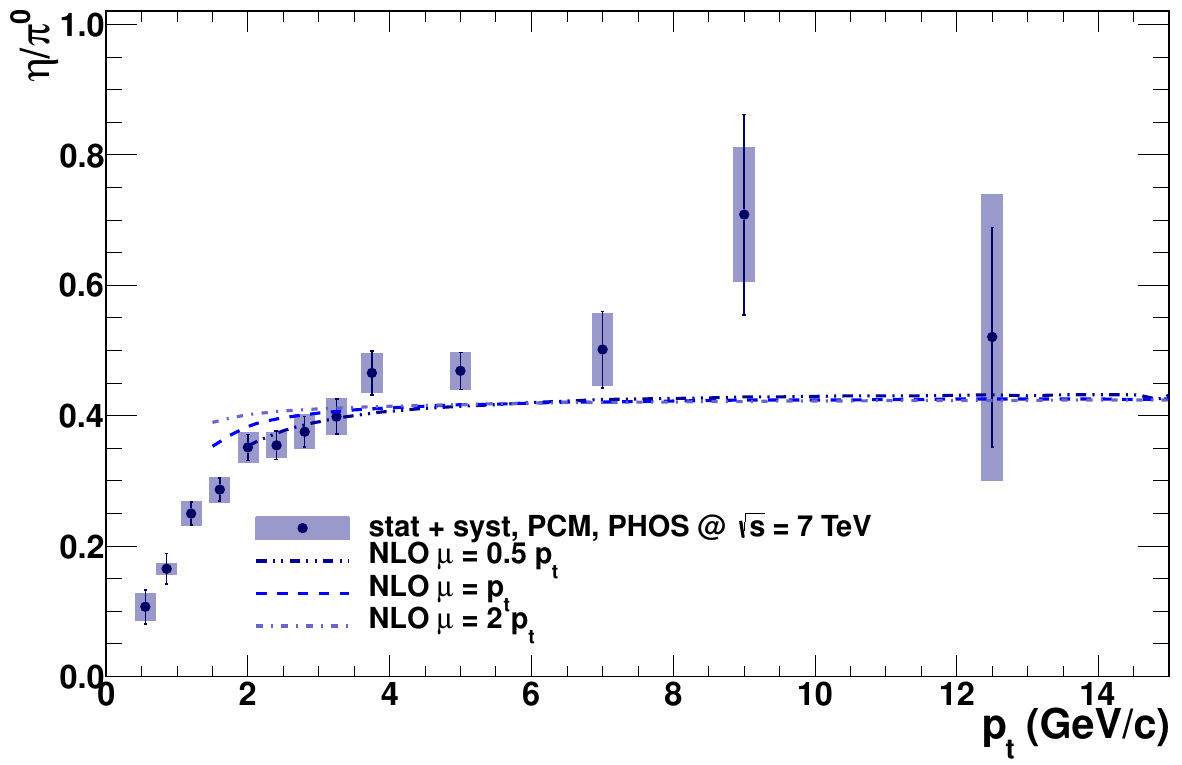}
  \caption{$\eta/\pi^0$ ratio measured in \pp ~collisions at
    $\sqrt{s}=7$~TeV compared to NLO pQCD predictions.}
  \label{fig:EtaToPi0}
\end{figure}

The ratio $\eta/\pi^0$ is shown in Fig.~\ref{fig:EtaToPi0}. It has the
advantage that systematic uncertainties in the measurement partially
cancel. This is also the case for the NLO pQCD calculation, where in particular the
influence of the PDF is reduced in the ratio.  Here,
predictions that failed to reproduce the measured $\pi^0$ and $\eta$ cross
section are able to reproduce the $\eta/\pi^0$ ratio.

\section{Conclusion}

In summary, the invariant differential cross sections for inclusive
$\pi^0$ production in \pp\ collisions at $\sqrt{s}=7$~TeV and 0.9~TeV and for $\eta$
meson production at 7~TeV have been measured in a wide $\pT$ range
taking advantage of two independent methods available in the ALICE
experiment at the LHC. NLO pQCD calculations cannot provide a
consistent description of measured data at both beam
energies. State-of-the-art calculations describe the data at 0.9 TeV
and 0.2 TeV \cite{Adare:2007dg},
however this is not the case at
7~TeV, where the calculations overestimate the cross sections and exhibit
a different slope compared to the data. Thus, this measurement provides an
important input for the tuning of pQCD calculations and represents
crucial reference data for the measurement of the nuclear modification
factor $R_{\rm AA}$ of the $\pi^0$ production in heavy-ion collisions at
the LHC.  
Furthermore, the NLO predictions for the $\eta$ mesons using the
newest fragmentation functions require a value $\mu=2\pT$ in order to get 
closer to the experimental results.

\section{Acknowledgments}

The ALICE collaboration would like to thank all its engineers and technicians for their invaluable contributions to the construction of the experiment and the CERN accelerator teams for the outstanding performance of the LHC complex.
\\
The ALICE collaboration acknowledges the following funding agencies for their support in building and
running the ALICE detector:
 \\
Calouste Gulbenkian Foundation from Lisbon and Swiss Fonds Kidagan, Armenia;
 \\
Conselho Nacional de Desenvolvimento Cient\'{\i}fico e Tecnol\'{o}gico (CNPq), Financiadora de Estudos e Projetos (FINEP),
Funda\c{c}\~{a}o de Amparo \`{a} Pesquisa do Estado de S\~{a}o Paulo (FAPESP);
 \\
National Natural Science Foundation of China (NSFC), the Chinese Ministry of Education (CMOE)
and the Ministry of Science and Technology of China (MSTC);
 \\
Ministry of Education and Youth of the Czech Republic;
 \\
Danish Natural Science Research Council, the Carlsberg Foundation and the Danish National Research Foundation;
 \\
The European Research Council under the European Community's Seventh Framework Programme;
 \\
Helsinki Institute of Physics and the Academy of Finland;
 \\
French CNRS-IN2P3, the `Region Pays de Loire', `Region Alsace', `Region Auvergne' and CEA, France;
 \\
German BMBF and the Helmholtz Association;
\\
General Secretariat for Research and Technology, Ministry of
Development, Greece;
\\
Hungarian OTKA and National Office for Research and Technology (NKTH);
 \\
Department of Atomic Energy and Department of Science and Technology of the Government of India;
 \\
Istituto Nazionale di Fisica Nucleare (INFN) of Italy;
 \\
MEXT Grant-in-Aid for Specially Promoted Research, Ja\-pan;
 \\
Joint Institute for Nuclear Research, Dubna;
 \\
National Research Foundation of Korea (NRF);
 \\
CONACYT, DGAPA, M\'{e}xico, ALFA-EC and the HELEN Program (High-Energy physics Latin-American--European Network);
 \\
Stichting voor Fundamenteel Onderzoek der Materie (FOM) and the Nederlandse Organisatie voor Wetenschappelijk Onderzoek (NWO), Netherlands;
 \\
Research Council of Norway (NFR);
 \\
Polish Ministry of Science and Higher Education;
 \\
National Authority for Scientific Research - NASR (Autoritatea Na\c{t}ional\u{a} pentru Cercetare \c{S}tiin\c{t}ific\u{a} - ANCS);
 \\
Federal Agency of Science of the Ministry of Education and Science of Russian Federation, International Science and
Technology Center, Russian Academy of Sciences, Russian Federal Agency of Atomic Energy, Russian Federal Agency for Science and Innovations and CERN-INTAS;
 \\
Ministry of Education of Slovakia;
 \\
Department of Science and Technology, South Africa;
 \\
CIEMAT, EELA, Ministerio de Educaci\'{o}n y Ciencia of Spain, Xunta de Galicia (Conseller\'{\i}a de Educaci\'{o}n),
CEA\-DEN, Cubaenerg\'{\i}a, Cuba, and IAEA (International Atomic Energy Agency);
 \\
Swedish Reseach Council (VR) and Knut $\&$ Alice Wallenberg Foundation (KAW);
 \\
Ukraine Ministry of Education and Science;
 \\
United Kingdom Science and Technology Facilities Council (STFC);
 \\
The United States Department of Energy, the United States National
Science Foundation, the State of Texas, and the State of Ohio.
\\
This job was supported partially by the grant RFBR~10-02-91052.
We would like to thank W.\ Vogelsang for providing the NLO pQCD
calculations used in this paper.


\onecolumn
\appendix
\section{The ALICE Collaboration}
\label{app:collab}

\begingroup
\small
\begin{flushleft}
B.~Abelev\Irefn{org1234}\And
A.~Abrahantes~Quintana\Irefn{org1197}\And
D.~Adamov\'{a}\Irefn{org1283}\And
A.M.~Adare\Irefn{org1260}\And
M.M.~Aggarwal\Irefn{org1157}\And
G.~Aglieri~Rinella\Irefn{org1192}\And
A.G.~Agocs\Irefn{org1143}\And
A.~Agostinelli\Irefn{org1132}\And
S.~Aguilar~Salazar\Irefn{org1247}\And
Z.~Ahammed\Irefn{org1225}
% ...
\renewcommand\labelenumi{\textsuperscript{\theenumi}~}
\section*{Affiliation notes}
\renewcommand\theenumi{\roman{enumi}}
\begin{Authlist}
\item \Adef{0}Deceased
\item \Adef{Dipartimento di Fisica dell'Universita, Udine, Italy}Also at: Dipartimento di Fisica dell'Universita, Udine, Italy
\item \Adef{M.V.Lomonosov Moscow State University, D.V.Skobeltsyn Institute of Nuclear Physics, Moscow, Russia}Also at: M.V.Lomonosov Moscow State University, D.V.Skobeltsyn Institute of Nuclear Physics, Moscow, Russia
\item \Adef{Institute of Nuclear Sciences, Belgrade, Serbia}Also at: "Vin\v{c}a" Institute of Nuclear Sciences, Belgrade, Serbia
\end{Authlist}
\section*{Collaboration Institutes}
\renewcommand\theenumi{\arabic{enumi}~}
\begin{Authlist}
\item \Idef{org1279}Benem\'{e}rita Universidad Aut\'{o}noma de Puebla, Puebla, Mexico
\item \Idef{org1220}Bogolyubov Institute for Theoretical Physics, Kiev, Ukraine
\item \Idef{org1262}Budker Institute for Nuclear Physics, Novosibirsk, Russia
\item \Idef{org1292}California Polytechnic State University, San Luis Obispo, California, United States
\item \Idef{org14939}Centre de Calcul de l'IN2P3, Villeurbanne, France
\item \Idef{org1197}Centro de Aplicaciones Tecnol\'{o}gicas y Desarrollo Nuclear (CEADEN), Havana, Cuba
\item \Idef{org1242}Centro de Investigaciones Energ\'{e}ticas Medioambientales y Tecnol\'{o}gicas (CIEMAT), Madrid, Spain
\item \Idef{org1244}Centro de Investigaci\'{o}n y de Estudios Avanzados (CINVESTAV), Mexico City and M\'{e}rida, Mexico
\item \Idef{org1335}Centro Fermi -- Centro Studi e Ricerche e Museo Storico della Fisica ``Enrico Fermi'', Rome, Italy
\item \Idef{org17347}Chicago State University, Chicago, United States
\item \Idef{org1118}China Institute of Atomic Energy, Beijing, China
\item \Idef{org1288}Commissariat \`{a} l'Energie Atomique, IRFU, Saclay, France
\item \Idef{org1294}Departamento de F\'{\i}sica de Part\'{\i}culas and IGFAE, Universidad de Santiago de Compostela, Santiago de Compostela, Spain
\item \Idef{org1106}Department of Physics Aligarh Muslim University, Aligarh, India
\item \Idef{org1121}Department of Physics and Technology, University of Bergen, Bergen, Norway
\item \Idef{org1162}Department of Physics, Ohio State University, Columbus, Ohio, United States
\item \Idef{org1300}Department of Physics, Sejong University, Seoul, South Korea
\item \Idef{org1268}Department of Physics, University of Oslo, Oslo, Norway
\item \Idef{org1132}Dipartimento di Fisica dell'Universit\`{a} and Sezione INFN, Bologna, Italy
\item \Idef{org1315}Dipartimento di Fisica dell'Universit\`{a} and Sezione INFN, Trieste, Italy
\item \Idef{org1145}Dipartimento di Fisica dell'Universit\`{a} and Sezione INFN, Cagliari, Italy
\item \Idef{org1270}Dipartimento di Fisica dell'Universit\`{a} and Sezione INFN, Padova, Italy
\item \Idef{org1285}Dipartimento di Fisica dell'Universit\`{a} `La Sapienza' and Sezione INFN, Rome, Italy
\item \Idef{org1154}Dipartimento di Fisica e Astronomia dell'Universit\`{a} and Sezione INFN, Catania, Italy
\item \Idef{org1290}Dipartimento di Fisica `E.R.~Caianiello' dell'Universit\`{a} and Gruppo Collegato INFN, Salerno, Italy
\item \Idef{org1312}Dipartimento di Fisica Sperimentale dell'Universit\`{a} and Sezione INFN, Turin, Italy
\item \Idef{org1103}Dipartimento di Scienze e Tecnologie Avanzate dell'Universit\`{a} del Piemonte Orientale and Gruppo Collegato INFN, Alessandria, Italy
\item \Idef{org1114}Dipartimento Interateneo di Fisica `M.~Merlin' and Sezione INFN, Bari, Italy
\item \Idef{org1237}Division of Experimental High Energy Physics, University of Lund, Lund, Sweden
\item \Idef{org1192}European Organization for Nuclear Research (CERN), Geneva, Switzerland
\item \Idef{org1227}Fachhochschule K\"{o}ln, K\"{o}ln, Germany
\item \Idef{org1122}Faculty of Engineering, Bergen University College, Bergen, Norway
\item \Idef{org1136}Faculty of Mathematics, Physics and Informatics, Comenius University, Bratislava, Slovakia
\item \Idef{org1274}Faculty of Nuclear Sciences and Physical Engineering, Czech Technical University in Prague, Prague, Czech Republic
\item \Idef{org1229}Faculty of Science, P.J.~\v{S}af\'{a}rik University, Ko\v{s}ice, Slovakia
\item \Idef{org1184}Frankfurt Institute for Advanced Studies, Johann Wolfgang Goethe-Universit\"{a}t Frankfurt, Frankfurt, Germany
\item \Idef{org1215}Gangneung-Wonju National University, Gangneung, South Korea
\item \Idef{org1212}Helsinki Institute of Physics (HIP) and University of Jyv\"{a}skyl\"{a}, Jyv\"{a}skyl\"{a}, Finland
\item \Idef{org1203}Hiroshima University, Hiroshima, Japan
\item \Idef{org1329}Hua-Zhong Normal University, Wuhan, China
\item \Idef{org1254}Indian Institute of Technology, Mumbai, India
\item \Idef{org1266}Institut de Physique Nucl\'{e}aire d'Orsay (IPNO), Universit\'{e} Paris-Sud, CNRS-IN2P3, Orsay, France
\item \Idef{org1277}Institute for High Energy Physics, Protvino, Russia
\item \Idef{org1249}Institute for Nuclear Research, Academy of Sciences, Moscow, Russia
\item \Idef{org1320}Nikhef, National Institute for Subatomic Physics and Institute for Subatomic Physics of Utrecht University, Utrecht, Netherlands
\item \Idef{org1250}Institute for Theoretical and Experimental Physics, Moscow, Russia
\item \Idef{org1230}Institute of Experimental Physics, Slovak Academy of Sciences, Ko\v{s}ice, Slovakia
\item \Idef{org1127}Institute of Physics, Bhubaneswar, India
\item \Idef{org1275}Institute of Physics, Academy of Sciences of the Czech Republic, Prague, Czech Republic
\item \Idef{org1139}Institute of Space Sciences (ISS), Bucharest, Romania
\item \Idef{org27399}Institut f\"{u}r Informatik, Johann Wolfgang Goethe-Universit\"{a}t Frankfurt, Frankfurt, Germany
\item \Idef{org1185}Institut f\"{u}r Kernphysik, Johann Wolfgang Goethe-Universit\"{a}t Frankfurt, Frankfurt, Germany
\item \Idef{org1177}Institut f\"{u}r Kernphysik, Technische Universit\"{a}t Darmstadt, Darmstadt, Germany
\item \Idef{org1256}Institut f\"{u}r Kernphysik, Westf\"{a}lische Wilhelms-Universit\"{a}t M\"{u}nster, M\"{u}nster, Germany
\item \Idef{org1246}Instituto de Ciencias Nucleares, Universidad Nacional Aut\'{o}noma de M\'{e}xico, Mexico City, Mexico
\item \Idef{org1247}Instituto de F\'{\i}sica, Universidad Nacional Aut\'{o}noma de M\'{e}xico, Mexico City, Mexico
\item \Idef{org23333}Institut of Theoretical Physics, University of Wroclaw
\item \Idef{org1308}Institut Pluridisciplinaire Hubert Curien (IPHC), Universit\'{e} de Strasbourg, CNRS-IN2P3, Strasbourg, France
\item \Idef{org1182}Joint Institute for Nuclear Research (JINR), Dubna, Russia
\item \Idef{org1143}KFKI Research Institute for Particle and Nuclear Physics, Hungarian Academy of Sciences, Budapest, Hungary
\item \Idef{org18995}Kharkiv Institute of Physics and Technology (KIPT), National Academy of Sciences of Ukraine (NASU), Kharkov, Ukraine
\item \Idef{org1199}Kirchhoff-Institut f\"{u}r Physik, Ruprecht-Karls-Universit\"{a}t Heidelberg, Heidelberg, Germany
\item \Idef{org20954}Korea Institute of Science and Technology Information
\item \Idef{org1160}Laboratoire de Physique Corpusculaire (LPC), Clermont Universit\'{e}, Universit\'{e} Blaise Pascal, CNRS--IN2P3, Clermont-Ferrand, France
\item \Idef{org1194}Laboratoire de Physique Subatomique et de Cosmologie (LPSC), Universit\'{e} Joseph Fourier, CNRS-IN2P3, Institut Polytechnique de Grenoble, Grenoble, France
\item \Idef{org1187}Laboratori Nazionali di Frascati, INFN, Frascati, Italy
\item \Idef{org1232}Laboratori Nazionali di Legnaro, INFN, Legnaro, Italy
\item \Idef{org1125}Lawrence Berkeley National Laboratory, Berkeley, California, United States
\item \Idef{org1234}Lawrence Livermore National Laboratory, Livermore, California, United States
\item \Idef{org1251}Moscow Engineering Physics Institute, Moscow, Russia
\item \Idef{org1140}National Institute for Physics and Nuclear Engineering, Bucharest, Romania
\item \Idef{org1165}Niels Bohr Institute, University of Copenhagen, Copenhagen, Denmark
\item \Idef{org1109}Nikhef, National Institute for Subatomic Physics, Amsterdam, Netherlands
\item \Idef{org1283}Nuclear Physics Institute, Academy of Sciences of the Czech Republic, \v{R}e\v{z} u Prahy, Czech Republic
\item \Idef{org1264}Oak Ridge National Laboratory, Oak Ridge, Tennessee, United States
\item \Idef{org1189}Petersburg Nuclear Physics Institute, Gatchina, Russia
\item \Idef{org1170}Physics Department, Creighton University, Omaha, Nebraska, United States
\item \Idef{org1157}Physics Department, Panjab University, Chandigarh, India
\item \Idef{org1112}Physics Department, University of Athens, Athens, Greece
\item \Idef{org1152}Physics Department, University of Cape Town, iThemba LABS, Cape Town, South Africa
\item \Idef{org1209}Physics Department, University of Jammu, Jammu, India
\item \Idef{org1207}Physics Department, University of Rajasthan, Jaipur, India
\item \Idef{org1200}Physikalisches Institut, Ruprecht-Karls-Universit\"{a}t Heidelberg, Heidelberg, Germany
\item \Idef{org1325}Purdue University, West Lafayette, Indiana, United States
\item \Idef{org1281}Pusan National University, Pusan, South Korea
\item \Idef{org1176}Research Division and ExtreMe Matter Institute EMMI, GSI Helmholtzzentrum f\"ur Schwerionenforschung, Darmstadt, Germany
\item \Idef{org1334}Rudjer Bo\v{s}kovi\'{c} Institute, Zagreb, Croatia
\item \Idef{org1298}Russian Federal Nuclear Center (VNIIEF), Sarov, Russia
\item \Idef{org1252}Russian Research Centre Kurchatov Institute, Moscow, Russia
\item \Idef{org1224}Saha Institute of Nuclear Physics, Kolkata, India
\item \Idef{org1130}School of Physics and Astronomy, University of Birmingham, Birmingham, United Kingdom
\item \Idef{org1338}Secci\'{o}n F\'{\i}sica, Departamento de Ciencias, Pontificia Universidad Cat\'{o}lica del Per\'{u}, Lima, Peru
\item \Idef{org1146}Sezione INFN, Cagliari, Italy
\item \Idef{org1115}Sezione INFN, Bari, Italy
\item \Idef{org1313}Sezione INFN, Turin, Italy
\item \Idef{org1133}Sezione INFN, Bologna, Italy
\item \Idef{org1155}Sezione INFN, Catania, Italy
\item \Idef{org1316}Sezione INFN, Trieste, Italy
\item \Idef{org1286}Sezione INFN, Rome, Italy
\item \Idef{org1271}Sezione INFN, Padova, Italy
\item \Idef{org1322}Soltan Institute for Nuclear Studies, Warsaw, Poland
\item \Idef{org1258}SUBATECH, Ecole des Mines de Nantes, Universit\'{e} de Nantes, CNRS-IN2P3, Nantes, France
\item \Idef{org1304}Technical University of Split FESB, Split, Croatia
\item \Idef{org1168}The Henryk Niewodniczanski Institute of Nuclear Physics, Polish Academy of Sciences, Cracow, Poland
\item \Idef{org17361}The University of Texas at Austin, Physics Department, Austin, TX, United States
\item \Idef{org1173}Universidad Aut\'{o}noma de Sinaloa, Culiac\'{a}n, Mexico
\item \Idef{org1296}Universidade de S\~{a}o Paulo (USP), S\~{a}o Paulo, Brazil
\item \Idef{org1149}Universidade Estadual de Campinas (UNICAMP), Campinas, Brazil
\item \Idef{org1239}Universit\'{e} de Lyon, Universit\'{e} Lyon 1, CNRS/IN2P3, IPN-Lyon, Villeurbanne, France
\item \Idef{org1205}University of Houston, Houston, Texas, United States
\item \Idef{org20371}University of Technology and Austrian Academy of Sciences, Vienna, Austria
\item \Idef{org1222}University of Tennessee, Knoxville, Tennessee, United States
\item \Idef{org1310}University of Tokyo, Tokyo, Japan
\item \Idef{org1318}University of Tsukuba, Tsukuba, Japan
\item \Idef{org21360}Eberhard Karls Universit\"{a}t T\"{u}bingen, T\"{u}bingen, Germany
\item \Idef{org1225}Variable Energy Cyclotron Centre, Kolkata, India
\item \Idef{org1306}V.~Fock Institute for Physics, St. Petersburg State University, St. Petersburg, Russia
\item \Idef{org1323}Warsaw University of Technology, Warsaw, Poland
\item \Idef{org1179}Wayne State University, Detroit, Michigan, United States
\item \Idef{org1260}Yale University, New Haven, Connecticut, United States
\item \Idef{org1332}Yerevan Physics Institute, Yerevan, Armenia
\item \Idef{org15649}Yildiz Technical University, Istanbul, Turkey
\item \Idef{org1301}Yonsei University, Seoul, South Korea
\item \Idef{org1327}Zentrum f\"{u}r Technologietransfer und Telekommunikation (ZTT), Fachhochschule Worms, Worms, Germany
\end{Authlist}
\endgroup


\end{document}